# Atom as a "Dressed" Nucleus

Vladimir Kalitvianski
Comissariat à l'Energie Atomique (CEA), Grenoble 38054, France
vladimir.kalitvianski@wanadoo.fr

We show that the electrostatic potential of an atomic nucleus "seen" by a fast charged projectile at short distances is quantum mechanically smeared due to nucleus motion around the atomic center of inertia. For example, the size of the "positive charge cloud" in the Hydrogen ground state is much larger than the proper proton size. For target atoms in excited initial states, the effect is even larger. The elastic scattering at large angles is generally weaker than the Rutherford scattering since the effective potential at short distances is softer than the Colombian one due to a natural "cutoff". In addition, the large-angle scattering leads to target atom excitations due to pushing the nucleus (=> inelastic processes). The Rutherford cross section is in fact inclusive rather than elastic. These results are analogous to those from QED. Non-relativistic atomic calculations are presented. The difference and the value of these calculations arise from nonperturbatively (exact) nucleus "dressing" that immediately leads to correct physical results and to significant technical simplifications. In these respects a nucleus bound in an atom is a simple but rather realistic model of a "dressed" charge in the QFT. This idea is briefly demonstrated on a real electron model (electronium) which is free from infinities.

**Keywords:** quantum mechanical charge smearing; natural cutoff of potential; interaction instead of self-action; non-perturbatively particle dressing; model of real electron.

PACS: 13.40.Gp, 14.60.Cd, 14.70.Bh, 11.10.Ef

## 1. INTRODUCTION

This paper resolves the old problem of self-action of elementary particles known from early Quantum Electrodynamics times. We show that the infinities in calculations appear not because of "pointlikeness" of the electron but mostly due to a very bad initial approximation used for interacting particles. The essence of the problem can be demonstrated in a simple atomic calculation. Then we explain how the correct theory of interacting particles can be formulated without self action and therefore without infinities related to this concept.

Sometimes one can read that it was hoped that quantum mechanics – a theory of wave functions – would somehow cure the problem related to the pointlike nature of the electron. The result, however, was disappointing [1]. Indeed, the self-interaction in QED remains, infinite corrections persist, and renormalization ideology leads to a rather bizarre notion of bare pointlike particles with infinite physical parameters.

The real particles are "dressed" or "renormalized". The bare particle perturbative "dressing" is awkwardly represented as a "vacuum polarization" effect due to creation of bare virtual particles which modifies the infinite, initial bare-particle potential at *long distances*. As a qualitative explanation of this "phenomenon", the Coulomb potential modification of the atomic nucleus at large distances due to electron screening is sometimes presented [2]:

"To draw an analogy in non relativistic quantum mechanics think of nuclei as bare atoms, electrons as virtual particles, atoms as dressed nuclei and the residual interaction between atoms, computed in the Born-Oppenheimer approximation, as the dressed interaction. Thus, for Argon atoms, the dressed interaction is something close to a Lennard-Jones potential, while the bare interaction is Coulomb repulsion. This is the situation physicists had in mind when they invented the notions of bare and dressed particles."

It would be a good analogy if the standard QFT calculations did not involve fictitious particles with infinite parameters (*i.e.* if the "bare" particles existed). But, as long as the standard QFT calculations involve infinities and renormalisations, the dressing physics remains quite vague and looks more like hand waving, even with the modern Wilson's approach.

On the other hand, there is a *much more realistic* (but practically unknown and thus unexploited) atomic analogy of particle dressing than that cited above. In this article we would like to bring it to your attention. Implementation of this idea in QED and in QFT removes the problems of appearing infinities.

### 1.1 Quantum Mechanical Charge Smearing

Everybody knows that the atomic electrons form a "negative charge cloud" within an atom. Few, however, know that a similar "cloud" is formed by the atomic nucleus around the atomic center of inertia. The "positive charge cloud" is just smaller in size – it is rescaled to the distances $r \leq a_0(m_e/M_A)$, but it is of exactly the same nature. Strictly speaking, a fast charged projectile capable of approaching the atomic center never meets the strong Coulomb repulsion there if it is scattered elastically. The nucleus coupling to the light atomic electrons naturally modifies the nucleus electrostatic potential *at short distances* $r \to 0$ which means its Coulomb singularity acquires a natural "cutoff". It is described in the frame of usual non-relativistic quantum mechanics and it is a real physical (observable) phenomenon. This radically corrects our understanding of "elementary" particle observation in a very well known example – the Rutherford scattering.

To bring it to light, we consider the simplest non-relativistic scattering of a fast $(v \approx 10v_0 \approx 0.07c)$, heavy structureless charge $Z_1$ with mass $M_1$ (a proton, for example) from a light atom with the nucleus charge $Z_A$. The projectile energy is then sufficient to test the atomic electrostatic potential at all short distances. With such velocities, no bound states between the projectile and the target may be formed, so we can safely speak of asymptotically "free" in- and out- atomic and projectile states (weak and finite-range interaction). This is a typical and a very old scattering problem in the Atomic Physics that can be considered quite accurately in the first Born approximation.

Usually it goes without saying that the nucleus stays in the atomic center of inertia (CI) and for large scattering angles the elastic cross section coincides with the Rutherford one [3]. At first sight this appears quite justified since the atomic electrons cannot repulse a heavy projectile backward; instead the Coulomb potential of the pointlike heavy nucleus seemingly comes into play. Unfortunately this explanation is inexact. The atomic CI is responsible for moving the atom as a whole and cannot describe the true effects of projectile-nucleus interaction, excitation of the atom when the transferred momentum is sufficiently big, for example. Assigning the Coulomb potential to the atomic CI also excludes the possibility of smearing the nucleus potential for an external observer. At the same time, considering the nucleus motion in the atom does not lead to any complications, at least in the scattering problem, but such a consideration is much more correct from physical point of view. It gives correct "second" (positive charge) atomic form factors that describe obvious and important physical effects. Now we will work out this simple problem in some details and point out close analogies with QED.

## 2. SECOND ATOMIC FORM FACTORS $f_n^{n'}(\mathbf{q})$

Let $\mathbf{r}_a$ be the electron coordinates relative to the atomic nucleus. The total atomic wave function is the product of the atomic CI plane wave and a wave function of the relative motion: $\Psi_A \propto \exp(i\mathbf{P}_{CI} \cdot \mathbf{R}_{CI}/\hbar)\psi_n(\mathbf{r}_a)$. And let $\mathbf{r}$ be the projectile coordinate (particle 1) relative to the atomic CI: $\mathbf{r} = \mathbf{r}_1 - \mathbf{R}_{CI}$. Then the microscopic potential of **electrostatic interaction** between the projectile and the atom is expressed as follows:

$$\hat{V} = Z_1 e^2 \left( Z_A \left| \mathbf{r} + \frac{m_e}{M_A} \sum_a \mathbf{r}_a \right|^{-1} - \sum_a \left| \mathbf{r} - \mathbf{r}_a + \frac{m_e}{M_A} \sum_{a'} \mathbf{r}_{a'} \right|^{-1} \right). \tag{1}$$

The differential cross section, calculated in the first Born approximation at the center of masses of the projectile and the target atom, is given by the formula [4]:

$$d\sigma_{n\ p}^{n'\ p'}(\theta) = \frac{4m^2 Z_1^2 e^4}{(\hbar q)^4} \frac{p'}{p} \left| Z_A \cdot f_n^{n'}(\mathbf{q}) - F_n^{n'}(\mathbf{q}) \right|^2 d\Omega. \tag{2}$$

This looks like the textbook formula but differs by the presence of the "second" or "positive charge" atomic form-factor $f_n^{n'}(\mathbf{q})$ which stands at the nucleus charge $Z_A$:

$$f_n^{n'}(\mathbf{q}) = \int \psi_{n'}^*(\mathbf{r}_a) \psi_n(\mathbf{r}_a) \exp\left( i \frac{m_e}{M_A} \mathbf{q} \sum_a \mathbf{r}_a \right) d\tau. \tag{3}$$

The second atomic form-factor is the effect of the nucleus binding to the atomic electrons. For elastic scattering ($n' = n$) it describes the "positive charge cloud" in the atom, while for $n' \neq n$ it gives the amplitude of atom exciting due to shaking the nucleus. There is a full analogy with the negative charge (electron) "cloud" and atom excitation amplitudes described by the first atomic form factor $F_n^{n'}$. The only difference is that the first and second atomic form factors "work" at quite different angles (or values of transferred momentum $\hbar\mathbf{q}$, or impact parameter regions). Our "first" or "negative charge" atomic form factor $F_n^{n'}$:

$$F_n^{n'}(\mathbf{q}) = \int \psi_{n'}^*(\mathbf{r}_a) \psi_n(\mathbf{r}_a) \left( \sum_a e^{-i\mathbf{q}\mathbf{r}_a} \right) \exp\left( i \frac{m_e}{M_A} \mathbf{q} \sum_a \mathbf{r}_a \right) d\tau \tag{4}$$

does not practically differ from the textbook one; it "works" at very small angles, and we will not need it anyway. We use the following (standard) notations for the center of mass variables and the atomic notation [3]:

$$m = \frac{M_1 \cdot M_A}{M_1 + M_A}, \quad \mathbf{p} = m\mathbf{v}; \quad \hbar\mathbf{q} = \mathbf{p}' - \mathbf{p}, \quad p' = \sqrt{p^2 - 2m(E_{n'} - E_n)}, \tag{5}$$

$$\hbar q = \sqrt{p'^2 + p^2 - 2pp'\cos\theta}, \quad \psi_n(\mathbf{r}_a) \equiv \psi_n(\mathbf{r}_1, \mathbf{r}_2, ..., \mathbf{r}_{Z_A}), \quad d\tau \equiv d^3 r_1 \cdot d^3 r_2 \cdot ... \cdot d^3 r_{Z_A},$$

$$a_0 = \hbar^2 / m_e e^2; \quad v_0 = e^2 / \hbar.$$

### 2.1. Elastic Scattering

Let us take a light atom in a quasi-stable initial state $\psi_n(\mathbf{r}_a)$ as a target. As can be seen from (3), the second atomic form-factor becomes essentially different from unity for elastic processes, $|f_n^{\ n}(\mathbf{q})| < 1$, when the scattering angle approaches or exceeds the value:

$$\theta_n = 2\arcsin\left\{\frac{v_0}{2v}\frac{a_0}{a_n}(1+\frac{M_A}{M_1})\right\}; \quad n \geq 0. \tag{6}$$

Then the elastic cross section becomes $|f_n^{\ n}|^{-2}$ times smaller than the Rutherford one.

Hereafter we will focus on the $\theta$ region $\theta_n \leq \theta \leq \pi$ and will refer to it as to "backward" scattering, regardless of the numerical values actually taken by $\theta$. (In fact, scattering to this angle range may be called "deep inelastic *atomic* scattering", in a full analogy with the deep inelastic scattering from hadrons. It might be used to study *atomic* structure at short distances; see the next section for more details). In this angle region the first atomic form factor $F_n^{\ n}(\mathbf{q})$ (due to the projectile-electron interaction) is negligible compared to the term $Z_A f_n^{\ n}(\mathbf{q})$ (determined with the projectile-nucleus interaction) so the projectile "feels" the atomic electron presence via the second atomic form factor rather than via the direct projectile-electron interaction. The physics is simple here: the electrons in atom make the nucleus move around the atomic CI and this smears the positive charge density via *quantum mechanical averaging* (7). It is a typical "vacuum field fluctuation" effect. As a result, the atom *elastically* repulses a positive projectile (or attracts a negative one) much weakly than the pointlike Coulomb center: (in the first Born approximation) the effective atomic electrostatic potential $U_n(\mathbf{r})$ "seen" with the projectile at short distances is equal to:

$$U_n(\mathbf{r}) = \int \psi_n^*(\mathbf{r}_a)\hat{V}(\mathbf{r},\mathbf{r}_a)\psi_n(\mathbf{r}_a)d\tau. \tag{7}$$

It does not grow to infinity as $1/r$ but remains finite when $r \to 0$. This effective potential may be considered as a "microscopic" one acting between a fast projectile and a non-elementary target (though this is useful only in the elastic backward scattering description):

$$d\sigma_{n\ p}^{n\ p}(\mathbf{q}) = \frac{m^2}{4\pi^2\hbar^4}\left|\int U_n(\mathbf{r})e^{-i\mathbf{qr}}d^3r\right|^2 d\Omega. \tag{8}$$

To simplify numerical illustration and to avoid consideration of identical particles we present a particular case of proton scattering from deuterium. The proton velocity is chosen to be $v = 10v_0$. Then $\theta_n \approx 0.3/(1+n)^2$; $n \geq 0$ and $f_0^{\ 0}(q(\theta)) = \left[1 + (100/9)\sin^2(\theta/2)\right]^{-2}$.

For proton-deuterium collisions the effective atomic potential (7) is shown in Fig. 1. The potential is compared to the Coulomb curve $e^2/r$ (dot-dashed line) and to a simple analytical approximation $U_0(r)_{appr.} = e^2/\sqrt{r^2 + (m_e/M_A)^2 a_0^2}$ (dot line). The distance $(m_e/M_A)a_0$ of the Coulomb "singularity" effectively "cutting off" appears here quite naturally thanks to the electron presence being *exactly* (rather than perturbatively) taken into account in (1), (7) and (8).

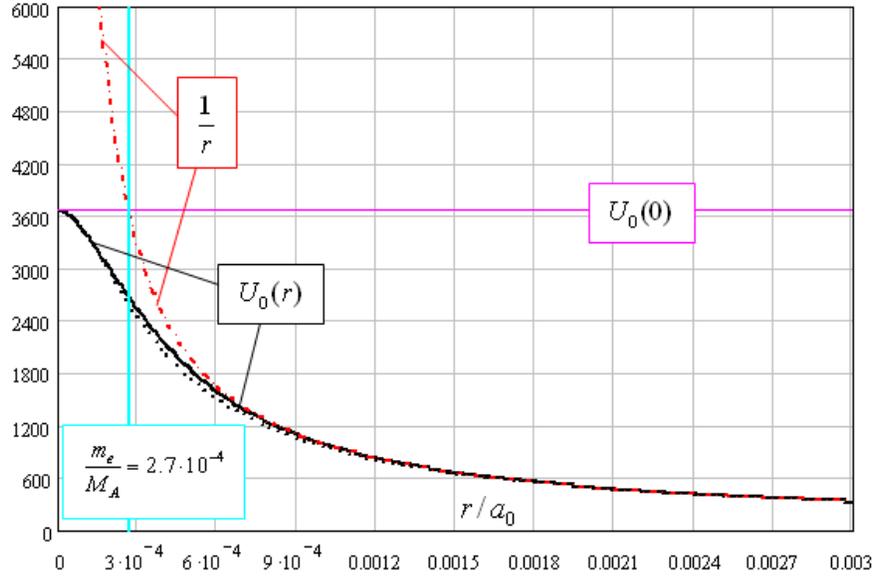

Fig. 1. The effective deuterium potential $U_0(r)$ "seen" by a fast proton as a function of distance to the atomic CI. At short distances $0 \leq r \leq (m_e/M_A)a_0$ it is essentially "softer" than the Coulomb one.

An attempt to fulfil a "perturbative" calculation of this amplitude, for example, with $(U_0)_{appr.} \propto r^{-1} - (m_e/M_A)^2 a_0^2 r^{-3}/2 + ...$ in (8) leads to corrections which are divergent at small distances, for example, $\int_0 (r^2 dr)/r^3 = (\ln r|_{r \to 0}) \to \infty$. As we can conclude from Fig. 1, the Coulomb potential $1/r$ is "infinitely far" from the exact effective potential $U_0(r)$ at short distances. The other Taylor terms are also distant there – the series in powers of $(m_e/M_A)(a_0/r)$ diverges when $r \to 0$. The corresponding integrals in (8) diverge too.

Of course, in these divergences there is no physics such as "vacuum polarization" due to "virtual" electron contributions. Rather, there is simply a very bad initial approximation of $U_0(r)$ (i.e. $\propto 1/r$) and therefore divergent iterative terms to "correct" it. Now it is clear why trying to calculate the smearing effects "perturbatively", i.e., by using $1/r$ as the initial approximation of interaction potential is not a good idea.

In the proton-deuterium example the maximum value of $U_0(\mathbf{r})$ is much smaller than the initial projectile kinetic energy: $U_0(0) = (e^2/a_0)(2m_p/m_e) = \alpha^2 2m_p c^2$, $mv^2/2 = (50/3)\alpha^2 2m_p c^2 \approx 17 \cdot U_0(0)$, so the proton can approach and "pass through" the positive cloud without problem. This fact also validates the applicability of the first Born approximation. It is obvious that accounting for the higher-order Born terms and spin cannot invalidate the smearing physics outlined above because it is nearly exact.

The positive charge cloud in an atom is rather similar to the negative (electron) charge cloud. For the Hydrogen atom with $a_0 \approx 0.53 \cdot 10^{-10}$ m the positive cloud has a size of about $2(m_e/M_p)a_0 \approx 5.8 \cdot 10^{-14}$ m. It is much smaller than the atomic size $2a_0$ but is still bigger than the proper proton size ($\leq 1.7 \cdot 10^{-15}$ m) determined with the Hofstadter's form factor.

The *most* important thing to note here is that even if the atomic nucleus and the projectile were structureless (just as in our simple calculation), their interaction potential (1) would be anyway effectively (quantum mechanically) cut off at small distances due to nucleus coupling to the atomic electrons (7). We have absolutely no need to seek or introduce any other (alien) mechanism of cutoff if we account for this one correctly (i.e., in the first turn).

The textbooks, which neglect the term $(m_e/M_A)\sum_a \mathbf{r}_a$ in the projectile-nucleus potential (1) (it is the distance from the atomic CI to the nucleus), erroneously substitute the nucleus coordinates with those of the atomic CI and give a physically wrong picture of elastic scattering: they obtain an unaltered Rutherford formula (no smearing effect is accounted). Thus the atomic nucleus is taught to be pointlike. QED similarly teaches that the electron is pointlike.

The curve $d\sigma_n^{\ n}(\theta) \propto \left|\int U_n(\mathbf{r})e^{-i\mathbf{qr}}d^3r\right|^2$, considered figuratively hereafter as the state $\psi_n$ "photograph", is rather "pale" and distorted by the factor $\left|f_n^{\ n}\right|^2$ in comparison with the Rutherford "picture" $d\sigma_{Ruth.}(\theta) \propto [\sin(\theta/2)]^{-4}$. In particular $\left|f_0^{\ 0}(q(\pi))\right|^2 \approx 4.7 \cdot 10^{-5} \ll 1$. Let us note here that in terms of transferred momentum $\hbar\mathbf{q}$ the positive-charge elastic atomic form-factor serves as a natural regularization factor (*momentum cutoff*) in the momentum space since it makes the elastic (Coulomb) amplitude $\propto 1/(\hbar\mathbf{q})^2$ tend rapidly to zero at big transferred momenta.

### 2.2. Inelastic Scattering

As one can see from (3), the second atomic form-factor is essentially different from zero for inelastic processes, $f_n^{\ n'}(\mathbf{q}) \neq \delta_{nn'}$, when the scattering angle approaches or exceeds the value $\theta_n$ (6). The physics is simple here – when the projectile transfers sufficiently big momentum to the nucleus, the relative motion of the atomic electrons and the nucleus in gets perturbed. This gives rise to excited final atomic states that is quite natural. This is similar to the atom exciting by "shaking" the electrons under small angle scattering. With the nucleus it just happens at much larger angles $\theta$. In this angle region the first inelastic atomic form factor $F_n^{\ n'}(\mathbf{q})$ is quite negligible compared to the term $Z_A f_n^{\ n'}(\mathbf{q})$.

No excitation can be obtained though ($f_n^{\ n'}(\mathbf{q}) \Rightarrow \delta_{nn'} = 0$) if one substitutes (unnecessarily and erroneously) the nucleus coordinates with the coordinates of the atomic CI. If the projectile "hits" the atomic center of inertia ($V(\mathbf{r}) \propto 1/r$), then the atom is only accelerated as *a whole* (bodily) *whatever momentum is transferred*. That is physically wrong.

Fig. 2 represents some inelastic second atomic form-factors $f_0^{\ n'}$. The corresponding inelastic cross sections are proportional to their squares.

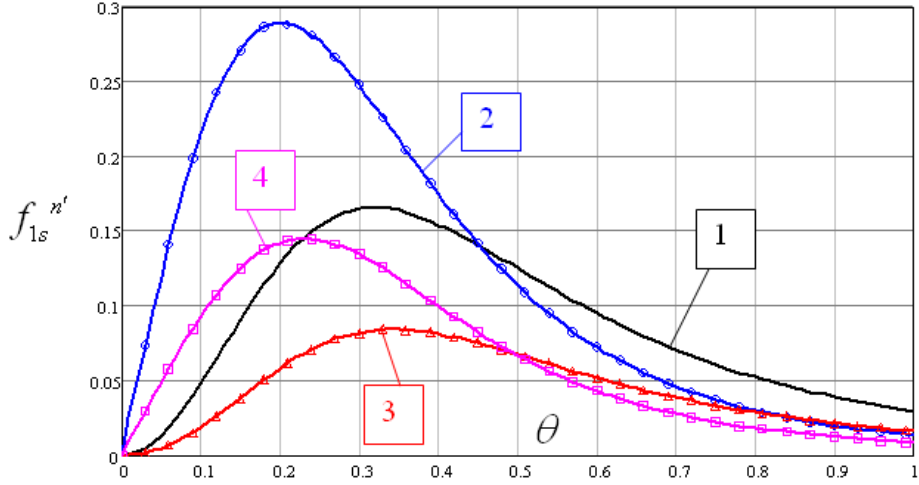

Fig. 2. $\theta$-dependence of $f_0^{n'}$ for exciting the following $|n',l',m'\rangle$ states of deuterium by fast ($v = 10 \cdot v_0$) proton: 1 - $|1,0,0\rangle$, 2 - $|1,1,0\rangle$, 3 - $|2,0,0\rangle$, 4 - $|2,1,0\rangle$. $\theta_0 = 0.3$.

The excitation cross sections $d\sigma_n^{n'}(\theta_n \leq \theta \leq \pi)$ can be measured experimentally. The projectile kinetic energy at these velocities is about several MeV. In practice there is no possibility of resolving the lost energy of such rapid projectiles with precision of order 10 – 100 eV. It is not even possible to prepare the incident beam with that level of energy accuracy. That is why dealing only with scattered projectiles inevitably gives the *inclusive* cross section.

Another matter is observation of recoil atoms. The excited atoms radiate. Atoms excited due to hitting electrons (described with $F_n^{n'}$ under small angle scattering) receive small momenta and radiate the standard spectral lines. Target atoms excited due to shaking the nucleus (determined with $f_n^{n'}$ under large angle scattering) receive bigger momenta, therefore their spectral lines will be somewhat shifted due to the Doppler effect. Registering simultaneously the scattered "backward" projectile *and* the shifted spectral lines permits the observer to distinguish different inelastic processes. Thus it is possible, in principle, to measure the elastic and different inelastic cross sections separately (2). For that the target atoms should obviously be in a cold, low-density gas state in order not to damp such excitations by the inter-atomic collisions.

### 2.3. The Inclusive Cross Section

If experimentally is counted only the *number* of scattered "backward" projectiles, without observing the target excitations (as Rutherford and many others did), what is measured is, in fact, the sum of elastic and inelastic cross sections. In this case the quantum mechanical result is very close to the Rutherford formula:

$$\frac{d\sigma_{incl}}{d\Omega} = \sum_{n'} \frac{d\sigma_{n\ p}^{n'\ p'}}{d\Omega} = \sum_{n'} \frac{4m^2 Z_1^2 Z_A^2 e^4}{(\hbar q)^4} \frac{p'}{p} \left| f_n^{n'}(\mathbf{q}) \right|^2 \approx \qquad (9a)$$

$$\approx \left( \frac{Z_1 Z_A e^2}{2mv^2 \sin^2(\frac{\theta}{2})} \right)^2 ; \quad \theta >> \frac{m_e}{m} \frac{v_0}{v} \frac{a_0}{a_n} . \tag{9b}$$

This is easy to prove: as the energy losses on atomic excitation $E_{n'} - E_n$ are always much smaller than that spent on the whole atom acceleration, one can neglect the dependence of $p'$ and $q$ upon $n'$. Then the sum (9a) factorizes and reduces accurately enough to the product of the Rutherford cross section (9b) and unit due to the matrix sum rule [3]:

$$\sum_{n'} \left| f_n^{n'} \right|^2 = \left| f f^+ \right|_n^n = 1 . \tag{10}$$

(To obtain exactly unity in (10), one needs to sum over *all* final states $n'$, not just those permitted by the energy conservation law. We note, however, that the contribution of energetically forbidden final states ($n' >> 1$) is so small in our case that the sum rule (10) approximately holds.)

The physical sense of this result is simple: in calculations of the *number* of fast particles scattered "backward" (observations made with photographic film, for example), one can consider the notion of a "free" nucleus with its (Coulomb in our case) potential as a target, but one should *never* think that in such an experiment the target atoms do not get excited! Our theoretical inclusive result corresponds to the factually inclusive experimental data. This is the only correct approach to the scattering description.

### 2.4. Atom, Electron and Neutron as Projectiles

If the projectile is not elementary itself (for example, if it is another atom), then the cross section (3) will simply contain a product of two atomic form-factors, one per atom:

$$\frac{d\sigma_{n\ p}^{n'\ p'}(\theta)}{d\Omega} = \frac{4m^2 Z_1^2 Z_2^2 e^4}{(\hbar q)^4} \frac{p'}{p} \left| f1_{n1}^{n1'}(\mathbf{q}) \cdot f2_{n2}^{n2'}(\mathbf{q}) \right|^2 , \quad \theta >> \frac{m_e}{m} \frac{v_0}{v} \frac{a_0}{a_n} . \tag{11}$$

The fast electrons can also be used as projectiles. For non-relativistic fast electrons (with $v \approx 10 v_0$) the Rutherford scattering attenuation will arise, according to (6), for $n \geq 10$, i.e. for Rydberg target atoms. But when $v \to c$, the velocity in (6) will be replaced with $v/\sqrt{1-v^2/c^2}$. Therefore the effects outlined above for electron scattering from the hydrogen *ground* state $\psi_0$ may well be observed for the incident electron energies $\geq 3.5\ MeV$.

If one neglects the weak dependence of $p'$ and $q$ upon $n'$, then the cross section (2) breaks down into two factors – the Rutherford cross section (scattering from a "free" pointlike nucleus) and the probability $\left| f_n^{n'}(\mathbf{q}) \right|^2$ of exciting the atom following such a scattering event. This (conditional) probability coincides in the main with that found by A. Migdal in 1939 in the problem of atom exciting with neutron [3]. In our approach this probability is obtained immediately and automatically in the first Born approximation if one uses the true nucleus coordinates rather than coordinates of atomic CI in the microscopic short-range neutron-nucleus potential:

$$\hat{V}_{Neutron-Nucleus}(\mathbf{r}_{Neutron} - \mathbf{r}_{Nucleus}) = \hat{V}_{Neutron-Nucleus}(\mathbf{r} + \frac{m_e}{M_A}\sum_a \mathbf{r}_a) \approx A\delta(\mathbf{r} + \frac{m_e}{M_A}\sum_a \mathbf{r}_a), \quad (12)$$

$$d\sigma_{n\ p}^{n'\ p'} = \frac{m^2}{4\pi\hbar^4} \frac{p'}{p} |A|^2 \cdot |f_n^{n'}(\mathbf{q})|^2 \, d\Omega. \quad (13)$$

The first factor including $|A|^2$ (however inexact it is) defines in this approximation the cross section of transferring the momentum $\hbar\mathbf{q}$ from the incident neutron to the atomic nucleus, while the factor $|f_n^{n'}(\mathbf{q})|^2$ represents the searched probability of exciting the atom assuming that such a momentum transfer has happened (or not exciting for $n' = n$). Therefore, the effects of the nucleus binding to the electrons are the same for any other kind of projectiles as soon as they transfer the same (big) momentum to the target.

## 3. DRESSED NUCLEUS

Non-relativistic quantum mechanics describing the electrons and nuclei as interacting de Broglie waves, gives quite understandable and measurable results in comparison with classical mechanics. The microscopic Coulomb potential in quantum mechanics acts between these waves, not between pointlike classical particles. De Broglie waves may form observable stationary states $\psi_n$ which never manifest any pointlike structure if one does not make technically unnecessary and physically erroneous "simplifications". In particular, neither negative nor positive charges in an atom are pointlike in the experiment and in the *correct* theory. Summing up different events $n \rightarrow n'$ does not create, strictly speaking, an "objective" notion of some "free" pointlike particle.

So far it has been fairly easy to accept corrections to the elastic scattering picture due to the nucleus motion in an atom. The real surprise comes when one considers an excited atom as a target. Common sense tells us that the more weakly electrons are bound in the atom (they all are in very distant orbits, for example), the more weak is their influence on the "backward" elastic scattering from the atom. It seems that here one may safely *neglect* the electron-nucleus binding. In the limit of a highly excited atom (a Rydberg atom, for example, with the electrons at "infinity") the elastic cross section has to automatically reduce to the Rutherford one for all angles. Formulae (6) and (3) indicate, however, that this classical expectation is completely wrong: there is an even *stronger* attenuation of the elastically scattered backward flux from an excited atom – the positive cloud size actually increases with increasing $n$: $a_n \approx a_0(1+n)^2$; $n \geq 0$. The higher the value of $n$, the wider the positive (quantum mechanical) cloud, and the smaller $|f_n^n|^2$ at a given angle $\theta$ (Fig. 3). So when the target atom is "very big", the "backward" elastic scattering vanishes at practically all finite angles $\theta$ ($n \rightarrow \infty$, $\theta_n \rightarrow 0$, $|f_n^n| \rightarrow 0$) since the nucleus is smeared over the whole space. As the value of $U_n(0)$ decreases with $n \rightarrow \infty$, (Fig. 4) the validity of the first Born approximation increases. Therefore, one *cannot* prepare a pointlike nucleus just by "keeping" the atomic electrons "far away" in the initial and the final target states. This conclusion may seem highly anti-intuitive but it is a strict quantum mechanical result. In fact, there is no paradox here as the elastic process is simply *substituted* with inelastic ones: $\sum_{n' \neq n} |f_n^{n'}|^2 \rightarrow 1$. One *cannot* and indeed *need not* get rid of this effect in the correct theory. On the contrary, such a theory is much richer as it completely corresponds to the physical reality.

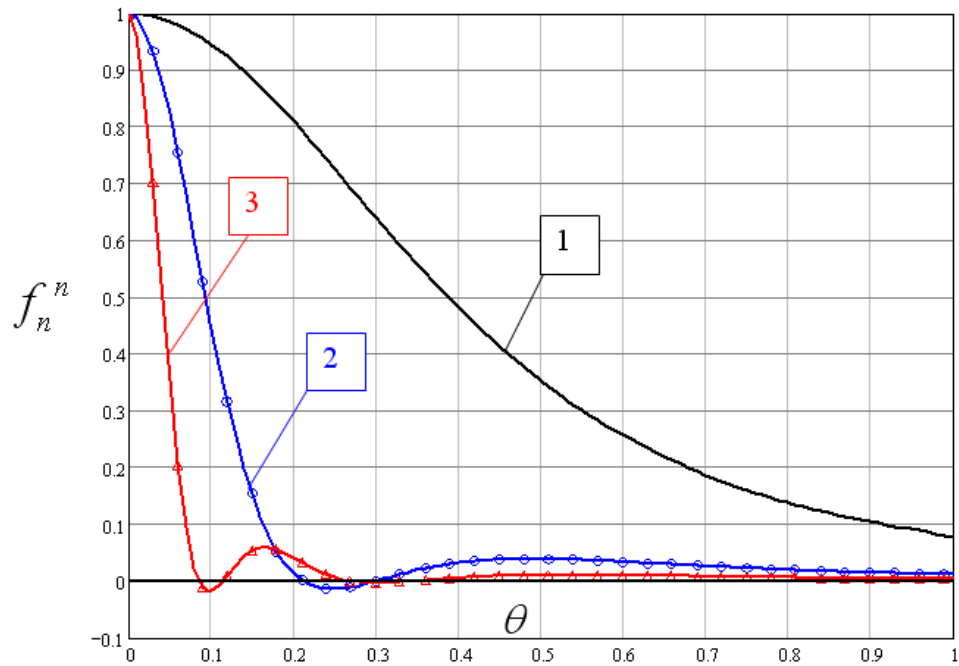

Fig. 3. $\theta$-dependence of $f_n^n$ for fast ($v = 10 \cdot v_0$) proton scattering from the following deuterium $|n,l,m\rangle$ states: 1 - $|0,0,0\rangle$, 2 - $|1,0,0\rangle$, 3 - $|2,0,0\rangle$. $\theta_0 \approx 0.3$; $\theta_1 \approx 0.08$; $\theta_2 \approx 0.03$.

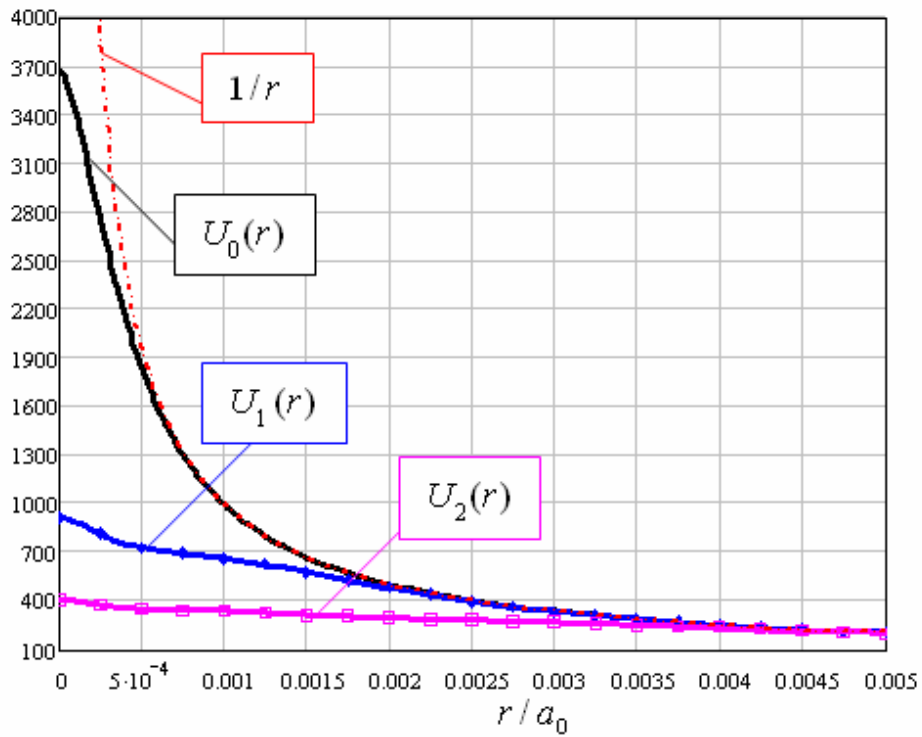

Fig. 4. The effective deuterium potentials $U_n(r)$ (6) "seen" by fast proton as a function of distance to the atomic CI. $n = 0, 1, 2$; $l = 0$; $m = 0$.

In actuality, any classical experimental result is *the inclusive picture* – the sum, like (9a), of compound quantum mechanical targets which were "broken" differently in the course of "observing".

Without good resolution, "poor" experiments pile up *different* events and produce an impression of observing some "objective" pointlike and elementary target (of course, with the help of our simplified notions of it). Thus, experimentally the classical picture (pointlike particle) is literally created as a cinematographic illusion obtained by *superimposing* (9a) all particular images of quite *different* elastic and inelastic events ("frames" of $d\sigma_n^{n'}(\theta)$) and voluntary assigning the inclusive result to *one "elementary"* particle.

As soon as we distinguish experimentally the elastic and inelastic processes (with atoms it is possible), we will never get the Rutherford cross section that always indicates *non-elementary* target structure. In other words, in nature there are no potential Coulomb singularities created by "free" pointlike *elementary* particles, and our theory (1)-(10) is in complete correspondence with this.

## 4. DRESSED ELECTRON (ELECTRONIUM)

### 4.1. Analogy with the scattering problem in QED

It is important to point out a fundamental physical analogy between the effects considered in this work, and those of quantum electrodynamics. In QED, strictly speaking, there is no elastic scattering either: any scattering is accompanied with some soft radiation. It is the inclusive cross section that reduces to the Rutherford one (or more generally, to the "mechanical" cross section) in QED [5-6]. Experimentally observing the bremsstrahlung in electrodynamics (quantum oscillator excitations) is physically analogous to observing the target atom excitations $n \to n'$ in "backward" scattering. In both cases the energy expended on excitations is much smaller than that spent on the whole target acceleration. Under these conditions pioneering experimentalists have dealt with the *inclusive* cross sections rather than with elastic ones. This fact explains why the notions of point-like elementary electrons and nuclei have appeared and are still so widespread.

In our atomic calculation the inelastic picture is obtained simply and naturally in the first Born approximation; this is so because we take into account the electrons' presence exactly. Their role is to provide the "vacuum fluctuation" effect (charge smearing for elastic processes) and to describe the natural inelastic channels for scattering from a non-elementary target. In QED such a result is also obtained but in higher orders and with a lot of technical complications (divergences of integrals, artificial regularizations, constant renormalizations, etc.). This is partially because the vacuum field fluctuations are considered *perturbatively* (in this case one also starts from the non smeared potential $1/r$ which is too far from reality), and partially because of the electron self-action terms.

Indeed, the electron coupling to the electromagnetic field has been proposed as the four-momentum "enlargement" in the free electron Hamiltonian:

$$p \to p + \frac{e}{c} A. \qquad (14)$$

This "minimal interaction" scheme works fine as long as the electromagnetic field is considered as external to the electron, but this kind of "coupling" leads to self-action if $A$ is the field radiated by the electron itself. Despite the non-physical exact solutions which arise in Classical Electrodynamics from such an ansatz (runaway solutions), this scheme was adopted in QED anyway. No wonder that the divergent self-energy terms reappear! In addition, considered perturbatively, the radiation term leads to the infrared catastrophe.

Is there any way to build the theory without self action, and to consider the radiation exactly rather than perturbatively? Yes, there is. Let us first consider the standard non-relativistic Hamiltonian in the frame of self-action ansatz:

$$H_{tot} = \frac{\mathbf{p}_1^2}{2M_1} + V(\mathbf{r}_1 - \mathbf{r}_e) + \left\{ \left(\mathbf{p}_e + \frac{e}{c}\mathbf{A}_{rad}\right)^2 / 2m_e + H_{osc} \right\}. \tag{15}$$

It may describe, for example, a scattering problem where the electron, permanently "coupled" to the quantized electromagnetic field, serves as a target for particle 1. The latter acts on the electron via potential $V(\mathbf{r}_1 - \mathbf{r}_e)$.

The exact solution accounting for the "influence" of the electromagnetic fluctuations on the electron has not been found, even for the "in" and "out" states. The perturbative treatment of the radiation field $\mathbf{A}_{rad}$ leads to physically incorrect results: in the first Born approximation the probability of not radiating any photon is equal to unity ($\Rightarrow$ elastic processes), whereas in reality this probability is equal to zero. In the next Born approximation the probability of emission of any photon diverges, etc.

To "cure" it in QED, in a full analogy with the atomic description, we have to look at the permanently coupled electron and the quantized electromagnetic field as at a compound system; let us call it an "*electronium*". Then, such a system has its own center of inertia with coordinates $\mathbf{R}_{CIe}$ and relative (internal) motion with coordinates describing the internal degrees of freedom (think of an atom as a model). The relative motion may receive energy if its state is perturbed. We know from experiments that "pushing" an electron transfers some energy into photon creation. It is the electromagnetic field oscillators which receive energy, so it is namely they that describe the relative ("internal") degrees of freedom of our electronium. In other words, the oscillator wave functions $\chi_{\mathbf{k},\lambda}(Q_{\mathbf{k},\lambda})$ play the same role for the electronium as the atomic wave function $\psi_n(\mathbf{r}_a)$ for an atom. The total *electronium* wave function is then the product of the electronium center-of-inertia plane wave and the oscillator wave functions:

$$\Psi_e \propto \exp(i\mathbf{P}_{CIe}\mathbf{R}_{CIe}/\hbar) \prod_{\mathbf{k},\lambda} \chi_{\mathbf{k},\lambda}(Q_{\mathbf{k},\lambda}). \tag{16}$$

The electronium non-relativistic Hamiltonian corresponding to such a solution is given with the formula:

$$H_e = (\mathbf{P}_{CIe})^2 / 2m_e + H_{osc}. \tag{17}$$

If (17) replaces the third term in (15), no self action is introduced and such a model leads to a correct physical description of the bremsstrahlung. So our ("*interaction*" instead of "self-action") ansatz is the following: instead of (15) we must use this:

$$H_{tot} = \frac{\mathbf{p}_1^2}{2M_1} + V(\mathbf{r}_1 - \mathbf{r}_e) + \left\{ \frac{\mathbf{P}_{CIe}^2}{2m_e} + H_{osc} \right\}. \tag{18}$$

The oscillator field tension $\mathbf{E}_{\mathbf{k},\lambda}$ is proportional to the oscillator canonical coordinate $Q_{\mathbf{k},\lambda}$ and the unit polarization vector $\mathbf{e}_{\mathbf{k},\lambda}$: $\mathbf{E}_{\mathbf{k},\lambda} \propto \mathbf{e}_{\mathbf{k},\lambda} \cdot Q_{\mathbf{k},\lambda}$. The projectile-electron coordinate $\mathbf{r}_1 - \mathbf{r}_e$ is expressed via the projectile-$CI_e$ coordinate $\mathbf{r} = \mathbf{r}_1 - \mathbf{R}_{CIe}$ and the oscillator fluctuating fields $\mathbf{E}_{\mathbf{k},\lambda}$ [7,8]:

$$\mathbf{r}_1 - \mathbf{r}_e = \mathbf{r}_1 - \mathbf{R}_{CIe} + \sum_{\mathbf{k},\lambda} \frac{-e \cdot \mathbf{E}_{\mathbf{k},\lambda}}{m_e c^2 \mathbf{k}^2} = \mathbf{r} + \sum_{\mathbf{k},\lambda} \frac{-e \cdot \mathbf{E}_{\mathbf{k},\lambda}}{m_e c^2 \mathbf{k}^2}. \tag{19}$$

The term $\sum_{\mathbf{k},\lambda} \frac{-e \cdot \mathbf{E}_{\mathbf{k},\lambda}}{m_e c^2 \mathbf{k}^2}$ for the electronium plays the same role as the term $\frac{m_e}{M_A}\sum_a \mathbf{r}_a$ for the atom (see (1)): it is the distance from the electronium CI to the electron. Keeping this term in (19) permits to act on the electron, while neglecting it means acting on the electronium CI, which leads to incorrect physics and to known mathematical problems in higher orders.

So, instead of "enlarging" the momentum $\mathbf{p}_e \to \mathbf{p}_e + (e/c)\mathbf{A}_{rad}$, we propose to "enlarge" the coordinates in the potential energy $\mathbf{r} \to \mathbf{r} + \sum_{\mathbf{k},\lambda} \frac{-e \cdot \mathbf{E}_{\mathbf{k},\lambda}}{m_e c^2 \mathbf{k}^2}$, to understand the kinetic energy and mass in (17) as the electronium CI energy and mass, and to understand the oscillator Hamiltonian as describing the relative (internal) electronium motion. By doing so, we take into account the quantized electromagnetic field in the "in" and "out" states exactly rather than perturbatively.

Then whatever the microscopic projectile-electron potential is (Coulomb or not), it does not contribute to the elastic cross section due to vanishing the elastic form-factor of the dressed electron (electronium), as it should be:

$$f_0^0(\mathbf{q}) = \int \exp\left(-i\mathbf{q}\sum_{\mathbf{k},\lambda}\mathbf{e}_{\mathbf{k},\lambda}\frac{e\cdot Q_{\mathbf{k},\lambda}}{m_e c^2 \mathbf{k}^2}\right) \prod_{\mathbf{k},\lambda} |\chi_{\mathbf{k},\lambda}|^2 dQ_{\mathbf{k},\lambda} \propto \exp\left(-\sum_{\mathbf{k},\lambda}(\mathbf{q}\mathbf{e}_{\mathbf{k},\lambda})^2 \frac{4\pi e^2}{m_e^2 k^3 c^3 V\hbar}\right)$$

$$= \exp\left(-\int_{k_{min}}^{k_{max}} (\mathbf{q}\mathbf{e})^2 \frac{4\pi e^2}{m_e^2 k^3 c^3 V\hbar} \frac{Vk^2 dk do}{(2\pi)^3}\right) = \exp\left(-e^2 \varsigma(q)\ln\frac{k_{max}}{k_{min}}\right)\bigg|_{k_{min}\to 0} \to 0. \tag{20}$$

That means quantum mechanical smearing of the electron charge over the whole space due to oscillator field fluctuations, so that the effective projectile-electronium elastic potential $U_0(\mathbf{r})$ (see (7)) is equal to zero for the elastic scattering.

It is easy to verify that all inelastic form-factors with finite number of final photons are also equal to zero, as it should be.

The totally inclusive cross section is different from zero and it is reduced accurately enough to the "mechanical" cross section [6] due to the sum rule (10), just as in the atomic case:

$$d\sigma_{incl}(\mathbf{q}) \approx \frac{m^2}{4\pi^2\hbar^4} \left|\int V(\mathbf{r})e^{-i\mathbf{qr}}d^3r\right|^2 d\Omega. \tag{21}$$

We see that the inclusive picture "corresponds" to scattering from the electronium CI as if the compound target were "pointlike", without internal degrees of freedom, and situated at the CI$_e$. Thus, when the quantized electromagnetic field is understood as the relative (internal) motion in a compound system (the electronium in our case), the scattering from such a system is automatically inelastic and inclusive in the first Born approximation, just as in the case of backward scattering from an atom as outlined above.

The interaction ansatz (16)-(19) naturally resolves the energy-momentum conservation laws for the bremsstrahlung (see formula (5)): one part of the projectile energy loss is spent on the target (electronium) acceleration as a whole and the remainder is spent on the target's internal energy increase (oscillator excitations). Now there is no need to neglect the electron recoil due to the radiation since in our model it is included automatically without problem.

No infrared divergences arise here since any dressed charge scattering becomes formally a *potential scattering of compound systems* with inevitable excitation of their "internal" (relative) degrees of freedom (photons), again, just as in the "backward" scattering of atoms where atomic electrons are "virtual". The obligatory inclusive consideration in such a theory yields the results corresponding to inclusive experiments. Thus, all classical results are obtained now due to *taking into account* the radiation processes (18)-(21), as in experiments, rather than due to neglecting them in (15) (i.e., the term $(e/c)\mathbf{A}_{rad}$). In other words, instead of saying that the soft radiation has a classical nature, it is correct to say that the classical radiation is the *inclusive* quantum mechanical result.

### 4.2. Bound electronium states

Hamiltonian (18) can also describe bound states of electronium and a nucleus (with particle 1 as a nucleus). Indeed, introducing the CI$_A$ and relative coordinates, we obtain:

$$H_{tot} = \frac{\mathbf{P}_{CI_A}^{2}}{2M_A} + \frac{\mathbf{p}^2}{2m} + V(\mathbf{r}) + \left[ V(\mathbf{r} + \sum_{\mathbf{k},\lambda} \frac{-e \cdot \mathbf{E}_{\mathbf{k},\lambda}}{m_e c^2 \mathbf{k}^2}) - V(\mathbf{r}) + H_{osc} \right]. \qquad (22)$$

The total (atomic) CI motion $\hat{\mathbf{P}}_{CI_A}^{2}/2M_A$ does not influence the bound-state spectrum, and we omit this part. The reminding Hamiltonian describes the relative motion of electronium and the nucleus (i.e., an atom). As we can see, it is not only the atomic potential $V(\mathbf{r})$ (which is the principal term) which takes part in creating the negative charge cloud in the atom, but also the oscillator potentials. In reality the oscillator numerical contributions are rather small. It is because in an atom the long-wave oscillators are forced to have the atomic frequency $\omega_0$ so their smearing effect is finite (see formula (20) with finite $k_{min}$). The easiest way to see it is to consider the operators in the Heisenberg representation or even the classical Hamilton equations $\dot{Q} = \partial H/\partial P$, $\dot{P} = -\partial H/\partial Q$: the oscillator field equations are coupled to the atomic variables due to the gradient (driving force) $\partial V(\mathbf{r} + \sum_{\mathbf{k},\lambda} \frac{-e \cdot \mathbf{E}_{\mathbf{k},\lambda}}{m_e c^2 \mathbf{k}^2})/\partial Q_{\mathbf{k},\lambda}$. For low-frequency oscillators this is the main force in comparison with the proper elastic force. That leads to the effective low-frequency cutoff in the oscillator spectrum, as if the whole spectrum had shifted $\omega_{\mathbf{k},\lambda} \to \omega_{\mathbf{k},\lambda} + \omega_0$. This fact then may be used in the usual Schrödinger picture to estimate the oscillator contributions. For that the exact interaction potential is expanded in powers of $\delta \mathbf{r} = -\sum_{\mathbf{k},\lambda} e \cdot \mathbf{E}_{\mathbf{k},\lambda}/m_e c^2 \mathbf{k}^2$:

$V(\mathbf{r} + \delta \mathbf{r}) \approx V(\mathbf{r}) + [\partial V(\mathbf{r})/\partial \mathbf{r}]\delta \mathbf{r} + (1/2)[\partial V(\mathbf{r})/\partial r_i \partial r_k]\delta r_i \delta r_k$. The approximate Hamiltonian

$\mathbf{p}^2/2m + V(\mathbf{r})$ gives a typical non-perturbed atomic spectrum. This spectrum is slightly corrected when the term $(1/2)[\partial V(\mathbf{r})/\partial r_i \partial r_k]\delta r_i \delta r_k$ is considered perturbatively. In particular, the average value of this term $\langle n|(1/2)[\partial V(\mathbf{r})/\partial r_i \partial r_k]\delta r_i \delta r_k|n\rangle$ (where $|n\rangle$ is the product of a non-perturbed atomic wave functions and oscillator wave functions with $\omega \geq \omega_0$), gives the atomic energy shifts known as the Lamb shifts [8] (the main non-relativistic part of it, of course). This is another validation that our ansatz (16)-(19) is the right approach to the QED formulation without infinities.

I think the relativistic Hamiltonian of Novel QED should be constructed in the same spirit. We propose to modify the usual Coulomb-gauge Hamiltonian in the following way: we must omit the term $\mathbf{j} \cdot \mathbf{A}_{rad}$ as originating from the wrong self-action ansatz (14). Instead we have to "insert" the oscillator variables into the electron coordinate $\mathbf{r}_e = \mathbf{R}_{CIe} - \sum_{\mathbf{k},\lambda} e g_{\mathbf{k},\lambda} \mathbf{E}_{\mathbf{k},\lambda}$ for each electron and positron so that any fermion would be constructed as an electronium – with *its own* oscillator Hamiltonian. Each "free" fermion Hamiltonian should be understood as the Hamiltonian of free-electronium CI motion. Any fermion-fermion distance in the four-fermion Coulomb interaction, say, $\mathbf{r}_1 - \mathbf{r}_2$ should be expressed via $CI_e$ and relative variables:

$$\mathbf{r}_1 - \mathbf{r}_2 = (\mathbf{R}_{CIe})_1 - (\mathbf{R}_{CIe})_2 - \left[\sum_{\mathbf{k},\lambda} e_1 (g_{\mathbf{k},\lambda} \mathbf{E}_{\mathbf{k},\lambda})_1 - \sum_{\mathbf{k},\lambda} e_2 (g_{\mathbf{k},\lambda} \mathbf{E}_{\mathbf{k},\lambda})_2\right], \quad \text{with}$$

$g_{\mathbf{k},\lambda} = \left(m_e c^2 \mathbf{k}^2 \sqrt{1 + (\hbar \mathbf{k}/m_e c)^2}\right)^{-1}$. This will provide each fermionium with its own form-factor due to its own oscillator field influence (as in (11)). Hence the relativistic Hamiltonian reads now:

$$H_{QED} = \int d^3 P \sum_{\substack{c=electron,\\positron}} \left\{\pi_c(\mathbf{P},t)\gamma^0(i\gamma\mathbf{P}\eta_c + m_e)u_c(\mathbf{P},t) + H_{osc.\mathbf{P},c}\right\} + \frac{1}{2}\int d^3 R_1 \int d^3 R_2 \frac{j^0(\mathbf{R}_1,t)j^0(\mathbf{R}_2,t)}{4\pi|\mathbf{r}_1 - \mathbf{r}_2|} \quad (23)$$

In the momentum space the electronium elastic form-factor, if non zero, serves as a *natural regularization factor* as it tends rapidly to zero when $|\mathbf{q}| \to \infty$ (useful in higher orders in relativistic calculations). So no ultraviolet divergences arise since first, no electron-radiation self-action is introduced in our electronium model, and second, the self-energy fermion loops originating in higher orders from the four-fermion Coulomb interaction vanish in scattering problems due to vanishing the elastic form-factors at each vertex. In bound states the electronium form-factor is non zero but it makes the loop contributions finite and small thanks to its significant regularization property. Thus, the problem of IR and UV divergences is removed in QED at one stroke by using the notion of an electronium (built in full analogy with the atom). No free nuclei exist in atoms; similarly no free electrons exist in nature. No bare constants are introduced, no renormalization is necessary. No connection between the "bare" and real charges appears in such a theory (there is no such a feature as the Landau pole, for example).

## 5. CONCLUSIONS

Even in non-relativistic quantum mechanics one can establish that a classical pointlike elementary particle is in fact nothing but the inclusive picture of many different and distinguishable (in principle) elastic and inelastic events. This understanding is much *deeper* than the usual quasi-classical limit $\hbar \to 0$ used in the proof of the correspondence principle.

Strictly speaking, an atom as a "dressed nucleus" does not manifest a pointlike (Coulomb) behavior at short distances $r \to 0$; neither does the real (dressed) electron which is always coupled to the quantized electromagnetic field. Taken into account correctly – in the first turn, the vacuum field fluctuations lead to the quantum-mechanical charge smearing and to the appearance of inelastic processes in the first Born approximation. No infinite "vacuum polarization" then arises to "screen" the "pointlike electron" field. This smearing physics cannot be obtained "perturbatively", even after renormalizations (see Section 2.1). That is why theorists have not invented anything more realistic than the "screening" (compensating) infinities or referring to unknown phenomena at the Plank scales.

Although the physics of the charge smearing outlined above is elementary, natural, and even known to some extent, its *fundamental* character has never been duly appreciated: the electron-field coupling is still considered as the self-action in QED. In this sense our "non perturbative" atomic calculations are rather instructive as, being flawless, they demonstrate how the correct physical theory can be constructed. If we accept the picture given in this article for a bound nucleus, then we are conceptually ready to admit the same picture for the real electron in QED – it is a compound system with a smeared quantum mechanically charge where the relative (or "internal") degrees of freedom are described with the photon oscillators. In other words, the radiated photons are just excited states of electronium.

We believe that the other "gauge" field theories should be reformulated in the same way: the corresponding self-action terms (gauge covariant derivative $D_\mu = \partial_\mu + eA_\mu$) should be replaced with the fermionium CI free motion derivative, the "gauge" field tensions should be "inserted" into the fermion coordinates to describe the relative (internal) degrees of freedom and symmetries of the corresponding compound "fermioniums". Then, for example, free quarks and gluons will not exist in the theory, in full agreement with non-existence of free electrons and photons in our electronium dynamics.